\documentclass{jfm}
\usepackage{ar}
\usepackage{float}
\usepackage{subfig}
\usepackage{wrapfig}
\usepackage{graphicx}
\begin{document}

\newtheorem{lemma}{Lemma}
\newtheorem{corollary}{Corollary}

\shorttitle{On the maintenance of an attached leading-edge vortex via model bird alula} %for header on odd pages
\shortauthor{T. Linehan and K. Mohseni} %for header on even pages

\title{On the maintenance of an attached leading-edge vortex via model bird alula}

\author{Thomas Linehan\aff{1}
  \corresp{\email{mohseni@ufl.edu}},
  Kamran Mohseni\aff{1}\aff{2}\aff{3}}

\affiliation{\aff{1}Department of Mechanical and Aerospace Engineering, University of Florida,
Gainesville, FL 32611, USA
\aff{2}Department of Electrical and Computer Engineering, University of Florida, Gainesville, FL 32611, USA
\aff{3}Institute for Networked Autonomous Systems, University of Florida, Gainesville, FL 32611, USA}

\maketitle

\begin{abstract}
Researchers have hypothesized that the post-stall lift benefit of bird's alular feathers, or alula, stems from the maintenance of an attached leading-edge vortex (LEV) over their thin-profiled, outer hand-wing. Here, we investigate the connection between the alula and LEV attachment via flow measurements in a wind tunnel. We show that a model alula, whose wetted area is 1\% that of the wing, stabilizes a recirculatory aft-tilted LEV on a steadily-translating unswept wing inclined at post-stall incidences. The attached vortex is the result of the alula's ability to smoothly merge otherwise separate leading- and side-edge vortical flows. We identify two key processes that facilitate this merging: i) the steering of spanwise vorticity generated at the wing's leading edge back to the wing plane and ii) an aft-located wall-jet of high-magnitude root-to-tip spanwise flow ($>$$80\%$ that of the freestream velocity). The latter feature induces LEV roll-up while the former feature tilts LEV vorticity aft and evacuates this flow toward the wing tip via an outboard vorticity flux. We identify the alula's streamwise position (relative to the leading-edge of the thin wing) as important for vortex steering and the alula's cant angle as important for high-magnitude spanwise flow generation. These findings advance our understanding of the likely ways bird's leverage LEVs to augment slow flight.
\end{abstract}

\section{Introduction}

 %-----------------------------
\begin{figure}
\begin{center}
\begin{minipage}{0.45\linewidth}  
\includegraphics[width=0.9\linewidth, angle=0]{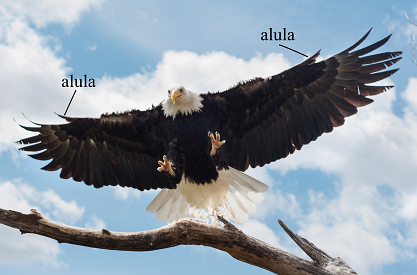}\\
\end{minipage}
 \end{center}
    \vspace{-10pt} 
\caption{\footnotesize Eagle landing with deflected alular feathers (or alula) as marked. }
   \label{fig:eagle}
    \vspace{-5pt} 
\end{figure}
%-----------------------------

The expedient landing ability of birds stems from a delicate maintenance of separated flows across their wings. In a glide-assisted landing, birds tilt their wings to high angles of attack to produce the necessary drag forces required to airbrake to safe touchdown speeds (\cite{VidelerJJ:05a,ThomasALR:07a}). Despite the leading-edge flow being separated, as evinced by the deflection of the lesser covert feathers (see Fig. \ref{fig:eagle}), birds sustain the ability to control attitude by maintaining flow reattachment over their outer wing through use of their alular feathers (\cite{ChoiH:15a}). However, precisely how the alular feathers accomplish this remains unclear and has been veiled in part by a lack of three-dimensional flow measurements of the alula in the literature. Here, we use Stereoscopic-Digital Particle Image Velocimetry (S-DPIV) to resolve the global, time-averaged, vortex flow over a canonical wing with model bird alulae and explain the mechanisms behind the alula's stall prevention ability. The following portion of the introduction reviews the literature relevant to this work. First, the historical treatment of the alula and its associated aerodynamics are reviewed. Then, the relevant literature regarding post-stall flows over finite wings are discussed. Lastly, prior work from the authors that is relevant to this work is reviewed and the current approach outlined.

\textit{The aerodynamics of the alula:}
A bird's alula consists of a cohort of feathers, approximately $15\%$ the length of the bird's wing (\cite{AlvarezJC:01a}), that stem from the bird's primary digit, or thumb (\cite{VidelerJJ:05a}). It was first observed on primitive birds living 115+ million years ago (\cite{SanzJL:96a,ZhouZ:00a,RauhutOWM:14a}) and is currently found on all modern bird species (minus hummingbirds) (\cite{VidelerJJ:05a}). 

The alula's function is largely aerodynamic, although research has indicated a possible sensory role (\cite{FeddeMR:93a}). During slow flight or flight requiring the bird's wing to be tilted at high angles, the alula increases wing lift by preventing stall (\cite{AlvarezJC:01a,VidelerJJ:05a,ThomasALR:07a,AndersonAM:07a,ChoiH:15a,WissaA:17a,WissaA:19a}) and has been shown to allow birds to perform steeper descents with greater changes in body orientation when landing (\cite{ChoiH:15a}). 

Despite consensus regarding the alula's stall-prevention ability, details of the underlying mechanisms remain unresolved. Aeroelastic observations of alula deflections during gliding portions of a bird's perching sequence show that the alula passively peels, then is actively protracted, from the plane of the wing (\cite{ThomasALR:07a}). The resulting gap formed between the alula and the top surface of the wing (see image in Fig. \ref{fig:eagle}) has resulted in early comparisons of the alula to flow control devices on aircraft such as leading-edge slots or slats that work by energizing the boundary layer to subvert its detachment and sustain wing lift (\cite{AlvarezJC:01a,SanzJL:05a,AndersonAM:07a}). However, this explanation does not account for the three-dimensionality of the alula, which, in its deflected state, mimics a miniature, canted, flap positioned at the leading-edge of the bird's wing.

The recent discovery of conical leading-edge vortices (LEVs) over the thin hand-wings of real (\cite{LentinkD:07a}) and model (\cite{VidelerJJ:04a}) swift wings has prompted a revaluation of the aerodynamics of the alula. \cite{VidelerJJ:04a} argued that the arm-wing and hand-wing of birds have differing lift-generating mechanisms; the thick-profiled arm-wing abides by conventional `attached-flow' aerodynamic principles and the thin hand-wings induce separated LEV flow. \cite{VidelerJJ:05a} suggests that the alula, which overhangs the thin hand-wing of birds, likely prevents stall via maintenance of separated edge flow rather than preventing flow separation from occurring in the first place.

Two updated interpretations of the aerodynamic function of the alula have been put forth (\cite{VidelerJJ:05a,ThomasALR:07a}). First, that the alula generates a small vortex which separates the attached-flow system on the arm-wing and the separated leading-edge vortex (LEV) on the hand-wing (\cite{VidelerJJ:05a}). This function has been partially corroborated by \cite{ChoiH:15a} due to planar PIV measurements of a streamwise alula vortex stemming from the alula and a stall-delaying effect outboard of the alula. Here, stall-delay on the outer wing was argued to be the result of root-to-tip spanwise flow induced by the streamwise-oriented alula vortex. However, the LEV was not resolved likely because flow measurements were two-dimensional and isolated to a region in the immediate vicinity of the alula. The second hypothesis is that the alula promotes LEV formation over the swept-back hand-wing of birds in flight scenarios when the arm-wing is completely stalled (\cite{VidelerJJ:05a,ThomasALR:07a}). Until recently, there has been no quantitative experimental evidence of the alula promoting LEV formation, nor a clear explanation on how it may do so. In this work, we will build upon our previous work elucidating the aerodynamics of the alula (described at the end of this section) and attempt to clarify the alula's role in LEV formation.

\textit{Post-stall flows over finite wings at high incidences:}
The leading-edge vortex (LEV) is a flow pattern leveraged by natural and man-made fliers to achieve high-lift generation during flight involving separated flow (\cite{EllingtonCP:96a, Dickinson:99a,ThomasALR:02a,WalkerSM:17a,SpeddingGR:08a,TobalskeBW:09a,VidelerJJ:04a,LentinkD:07a,TropeaC:10a}). Many interpretations of lift enhancement via LEV flow exist. The stationary vortex model of \cite{Saffman:77a} showed theoretically that a free vortex attached to the wing increases wing lift by inducing a stronger bound circulation around the wing. \cite{Mohseni:13ag} showed lift enhancement of the stablized LEV to be the combined effect of both LEV and TEV motion where an attached LEV, moving with the airfoil, decreases the negative lift contribution associated with LEV motion. Physically, the presence of a recirculatory LEV above the wing accelerates fluid on the wing's top surface which enhances lift via suction. In each interpretation, the key to harnessing LEV-lift is prolonging its residence near the wing surface for the time-scales relevant to the flight objective. Flapping fliers such as insects (\cite{EllingtonCP:96a, Dickinson:99a,ThomasALR:02a,WalkerSM:17a}) leverage complex kinematics to stabilize the LEV. Here, mechanisms such as Coriolis tilting (\cite{JardinT:17a,LentinkD:09a}), vorticity annihilation (\cite{BuchholzJHJ:14a}), and spanwise flow (\cite{DavidL:14a}) are suggested to stabilize the LEV and/or delay its detachment. Readers are referred to (\cite{EldredgeJD:19a}) for a detailed review of LEV mechanics on maneuvering aerofoils and wings. However, as this manuscript attempts to resolve the LEV-stabilizing ability of the alula in a glide-assisted landing scenario, the following review focuses on the fluid dynamics associated with steadily translating wings at post-stall incidence angles as opposed to maneuvering wings.

The fluid dynamics and associated loading of a finite wing at post-stall incidences is dominated by the growth, development, and interaction of the vortices stemming from the edges of the wing. The transient motion of the wing involves the roll-up of leading-, trailing-, and side-edge shear layers into a leading-edge vortex (LEV), trailing-edge vortex, and tip vortices, respectively, that collectively form an apparent vortex loop (\cite{FreymuthP:86a,Colonius:09b}).  \cite{Colonius:09b} showed via computations at $Re_c = O(100)$ that with increasing time the trailing-edge vortex is displaced further into the wake and the tip vortices form columnar structures. The LEV continues to accumulate spanwise vorticity, subsequently growing in length and eventually submits to shedding, likely when the length of the LEV approaches or exceeds that of the chord (\cite{RivalDE:14a}). Hair-pin-type leading-edge vortices are continually shed which disrupt the coherency of the columnar tip vortices. Around this time, the lift of the wing peaks and settles to a lower nominal value. At Reynolds numbers in the range $Re_c = O(10^4-10^5)$ the LEV and its shedding is transitional and is characterized by a Kelvin-Helmholtz instability in the leading-edge shear layer that results in the break up and shedding of discrete vortices that succumb to turbulence in their downstream evolution (\cite{AzumaA:11a,LiburdyJ:09a}). 

Regulating LEV growth and/or shedding is thus key to sustained lift generation during steady flight at post-stall incidences. Certain modifications to the geometry of the wing are known to achieve this. Wings of sufficiently low aspect ratio produce downward-induced velocity via tip vortex downwash that pin the LEV close to the wing plane (\cite{Colonius:09b}). This downwash forces LEV vorticity to recirculate above the wing restricting large-scale vortex shedding (\cite{Mohseni:17o}). \cite{Mohseni:17c} explains that tip vortex downwash limits the chord-wise growth of the LEV as indicated by the saddle point associated with the LEV lying on the wing surface as opposed to in the wake. The existence of the saddle point upstream of the trailing-edge results in a smooth merging of flow at the trailing-edge, reduced trailing-edge vorticity generation, and in consequence, sustained wing lift.

In addition to tip vortex downwash, LEV shedding can be delayed or curtailed at high incidences by encouraging the spanwise transport, or release, of vorticity toward the wing tip through use of curved or swept leading-edges (\cite{Colonius:09b}). On rectangular wings, the LEV and tip vortices remain separate structures due to the sharp corners of the wing, however, for the elliptical-type and semicircular test cases considered by \cite{Colonius:09b} there are no discontinuities in the vortex sheet that emanates from the leading-edge and wing tips. Moreover, the elimination of wing tips altogether through use of a delta-wing planform is readily understood to form two separate but stable LEVs on the left and right wing. 

\begin{figure}
\begin{center}
\begin{minipage}{0.32\linewidth}  
\includegraphics[width=0.9\linewidth, angle=0]{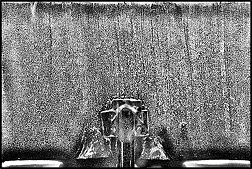}\\ \centering \footnotesize (a) baseline, no alula
\end{minipage}
\begin{minipage}{0.32\linewidth}  
 \includegraphics[width=0.9\linewidth, angle=0]{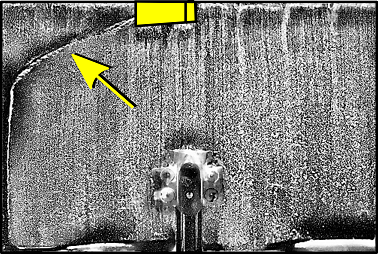}\\ \centering \footnotesize (b) leftward-oriented alula
\end{minipage}
\begin{minipage}{0.32\linewidth}  
  \includegraphics[width=0.9\linewidth, angle=0]{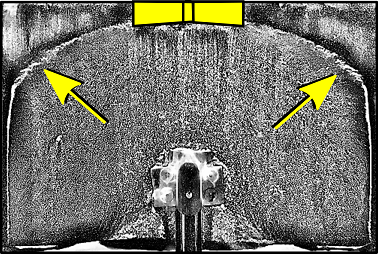}\\ \centering \footnotesize (c) dual-opposing alula
\end{minipage}
 \end{center}
    \vspace{-5pt} 
\caption{\footnotesize Surface-oil flow visualizations of an \AR\ = 1.5 wing at $\alpha = 25^\circ$ with (a) no alula, (b) leftward-oriented alula and (c) dual-opposing alula. Alula indicated by boxes. Separation line associated with an apparent vortex indicated by arrows. From \cite{Mohseni:19g}.}
   \label{fig:surfaceoil}
    \vspace{-5pt} 
\end{figure}

\textit{Prior work and current approach:}
The authors have been investigating the possibility that bird's alula can induce and stabilize a LEV on a wing in steady translation without requiring the wing to be swept back. This feature would enable the bird to maintain a spread-wing gliding posture that expedites landing through increased drag while subsequently utilizing LEV-lift to adjust attitude. In \cite{Mohseni:19g}, a canonical wing-alula configuration was considered where the wing and alula were modeled as thin, rectangular flat-plates with relative dimensions consistent with bird measurements. The alula was represented by a canted flap stationed at the leading-edge of the wing whose area was 1\% that of the wing. In this study, certain parameters of the wing and alula were varied, such as the angle of attack of the wing, the cant angle and spanwise position of the alula among others. The lift benefit of the alula was isolated to a range of post-stall incidences in the range 22 deg - 38 deg. In this range, surface-oil visualizations revealed the surface footprint of an apparent vortex that swept across the outer edges of the wing (see Fig. \ref{fig:surfaceoil}). The addition of a second alula resulted in two vortex systems on the left and right wing respectively. Lift and drag increment was found to be as high as $13\%$ and $25\%$ for certain single and dual alula configurations. In the current work, we use S-DPIV to elucidate these flow physics and connect this flow pattern to an aft-tilted LEV. Moreover, in the current work we explain the mechanisms of the alula that promote LEV formation and its attachment.

 %-----------------------------
%-----------------------------

The paper is organized as follows. The experimental set-up is described in Section \ref{sec:exp}. The global flow structure over the wing with and without alula(e) is described in Section \ref{sec:global}. The creation and stabilization mechanisms associated with the observed attached vortex system are investigated in Section \ref{sec:rollup}, Section \ref{sec:tilt}, and Section \ref{sec:spanflow} through analysis of LEV formation, vortex tilting, and outboard vorticity flux. In Section \ref{sec:carve} the reduced effectiveness of the alula at high angles of attack is considered. Concluding remarks regarding these findings are then given in Section \ref{sec:concl}.

\section{Experimental set-up}\label{sec:exp}

The wing planform and alula are similar to that described in \cite{Mohseni:19g}. The wing is an acrylic rectangular flat-plate with a chord of 9.53 cm and a span of 14.29 cm (\AR\ = 1.5). The rectangular planform is chosen to isolate the vortex stabilizing mechanisms of the alula from that associated with leading-edge curvature. The thickness of the wing was $0.047c$ where $c$ is the wing chord. All edges were left square.  Model alulae were mounted to the wing via 3D printed, low-profile, press-fit pins. The alula is represented as a rigid flat plate with a fixed geometry and orientation with its root positioned at the midspan of the wing. The alula is rectangular with a thickness of $0.008c$, a span of $0.15b$, and an area of $0.01S$ where $b$ is the wing length and $S$ is the wing area. Unless otherwise noted, the leading-edge of the alula is offset 0.007c in front of the leading-edge of the wing. The orientation of the deployed alula relative to the wing is defined by three angles: 1. The incidence angle defined by the angle of the alula's chord relative to the wing chord. 2. The deflection angle or cant angle, defined by the rotation of the alula from the plane of the wing. 3. The pronation angle, or the sweep angle of the alula (in the plane of the wing) relative to the wing's leading edge. The alula is canted off the plane of the wing at an angle of $25$ deg with an incidence angle and pronation angle fixed at zero degrees. During testing, the wing affixed with the alula is inclined to steady flow at an angle of attack of 28 deg with select cases repeated at an angle of attack of 36 deg. The flow speed was 12.1 m/s corresponding to a Reynolds number of 75,000 which is within the range of bird flight (\cite{VidelerJJ:04a}).

Two modified alula's were also tested: a `forward-shifted' alula, and an `overhung alula'. The forward-shifted alula had identical properties as the baseline alula (described above), however, it was shifted forward on the wing such that its mid chord was coincident with the wing's leading-edge (offset distance of 0.044c). The overhung alula, maintained the same chordwise location on the wing as the baseline alula but had its root vertically offset from the wing plane and possessed no cant angle. The gap between the overhung alula and the wing's top surface was set so that its frontal area was approximately equal to that of the baseline (canted) alula. Moreover, the span of the overhung alula was kept the same as the projected span of the canted alula.

 Each alula was printed using a 3D Systems Projet 2500 multijet printer. The printer has a net build volume (XYZ) of 294 $\times$ 211 $\times$ 144 mm with a 800 $\times$ 900 $\times$ 790 DPI resolution with 32 $\mu m$ layers. Resolution before post processing is $\pm 0.025-0.05$ mm per 25.4 mm of part dimension. The material was VisiJet M2 RWT.

Wind tunnel experiments were performed in the Engineering Laboratory Design recirculating wind tunnel located at the University of Florida. The test section has a 61 $\times$ 61 cm$^2$ cross-section and is 2.44 m in length. The wind tunnel can achieve freestream velocities ranging from 3 - 91.4 m/s. The turbulence intensity of the freestream was 0.12\% at the tested velocity. 

\begin{figure}
\centering 
     \vspace{0pt}
  \begin{minipage}{0.8\linewidth}
   \includegraphics[width=1\textwidth, angle=0]{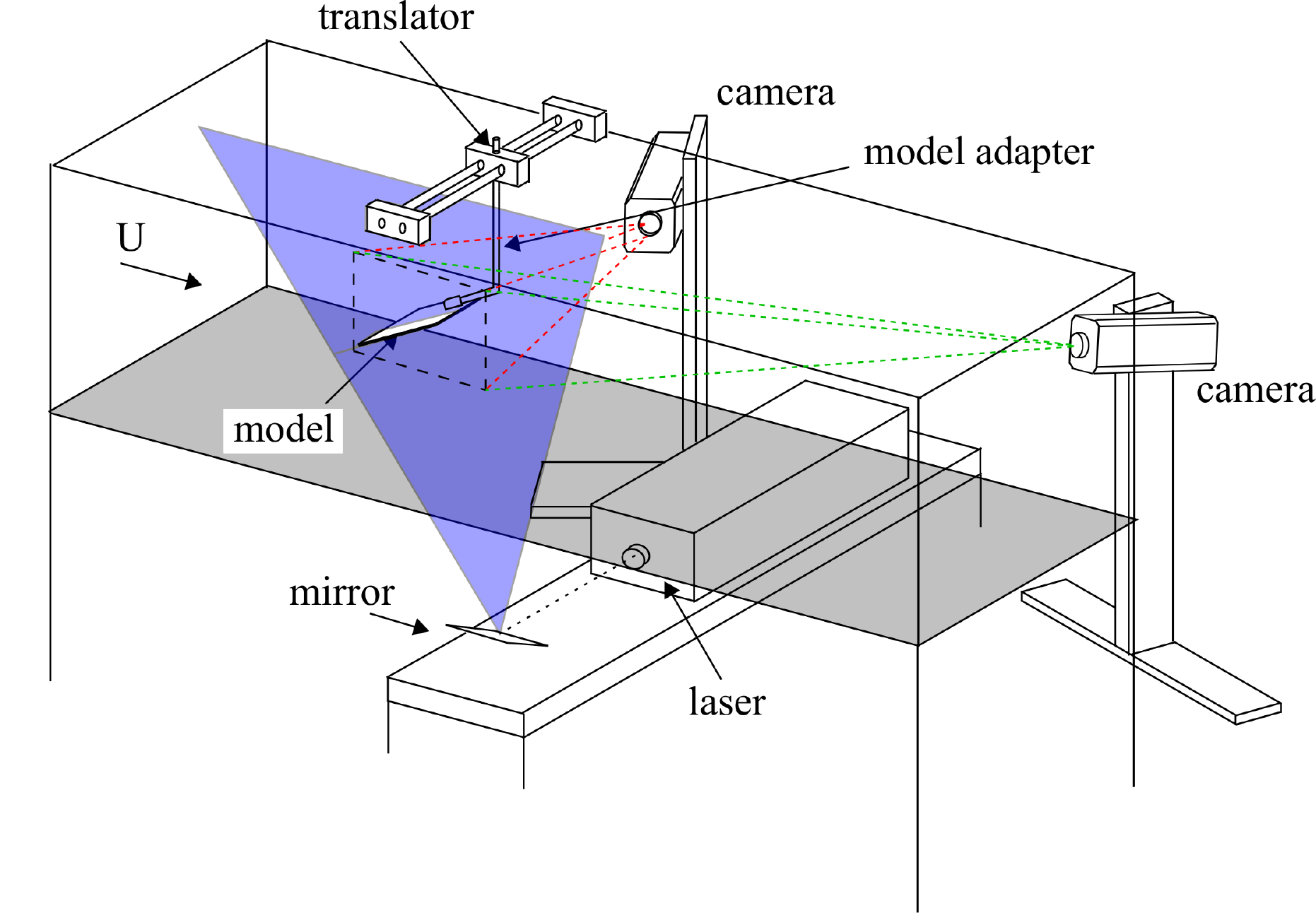}\\ 
   \end{minipage}     
        \vspace{-10pt}
       \caption{\footnotesize Stereo-Digital Particle Image Velocimetry (S-DPIV) system in wind tunnel.}
     \vspace{-10pt}
   \label{fig:PIVsch}
\end{figure}

A Stereo-Digital Particle Image Velocimetry (S-DPIV) system (see Fig. \ref{fig:PIVsch}) was used to measure the three-component velocity field in streamwise planes of the flow (2D-3C). The wind tunnel was seeded with $\sim$1 $\mu$m olive oil particles generated by an atomizer. These particles were illuminated by a 4 mm thick laser sheet generated by a 20 mJ Nd:YLF laser (Quantronix Darwin Duo, $\lambda=527$~nm). The imaging system consists of high-speed CMOS 1 Mpx cameras (Phantom v210/v211, 1280 $\times$ 800 px$^2$) with the object-to-image plane mapping function (\cite{AdrianRJ:97a}) determined with a precision-machined, dual-plane calibration target. Misalignment of the target with the laser sheet was corrected with the disparity map method (\cite{AdrianRJ:97a,WillertC:97a,WienekeB:05a}) for which 100 images (of the undisturbed freestream) were used.

  %-----------------------------
%----------------------------

An approximation of the mean volumetric flow field (3D-3C) is constructed from closely-spaced planes of data collected by translating the wing-alula model through a stationary vertically-oriented laser plane. 38 equally-spaced streamwise planes of data were taken a distance $0.06c$ apart on the wing where $c$ is the chord length of the wing. Each plane of S-DPIV data consists of 300 images taken at a rate of 100 Hz (3 sec of acquisition time). Each image was processed with Insight 4G software by TSI Inc. Images were first dewarped according to calibration images taken for each camera. Thereafter, an iterative multi-pass DPIV evaluation algorithm consisting of windowing shifting/deformation was performed on each image pair. Interrogation windows were made rectangular starting from 40 $\times$ 40 px$^2$ down to 20 $\times$ 20 px$^2$ (50\% overlap). The resulting spatial resolutions of the volumetric flow measurements in the horizontal, vertical, and streamwise directions are $\Delta y = 0.06c, \> \Delta z =  \Delta x = 0.026c $. The size of the measurement volume is 2.94c $\times$ 2.28c $\times$ 2.94c where the total number of measured velocity vectors is 113 $\times$ 38 $\times$ 75. Measured quantities such as circulation are computed using this grid. A refined grid with four times the resolution, i.e. 452 $\times$ 152 $\times$ 300, is used for three-dimensional plots. 

Statistics of S-DPIV measurements of the undisturbed freestream were used to quantify measurement errors. Taking each time-averaged velocity measurement in space as a single sample, velocity errors corresponding to twice the standard deviation of the sampling distribution were $e_{u}/U = 0.02$, $e_{v}/U= 0.01$, and $e_{w}/U = 0.02$. Vorticity is computed using the local circulation method \cite{Raffel:98a}. An estimate of the error in vorticity from this method is $e_{\omega} c/{U} = 0.61e_{U}c/(U\Delta x) = 0.72$ where $e_{U}$ is taken as the average of the above velocity errors. 

Unless otherwise noted, the coordinate system used to represent the data is in the laboratory axis where $x$ is downstream, $z$ is toward azimuth and $y$ is out the right wing as observed from an observer at the trailing-edge of the wing facing the leading-edge of the wing.

\section{Post-stall vortical flow over a wing with an alula}\label{sec:global} 
\begin{figure}
\centering
     \vspace{0pt}
  \begin{minipage}{1\linewidth}
   \includegraphics[width=1\textwidth, angle=0]{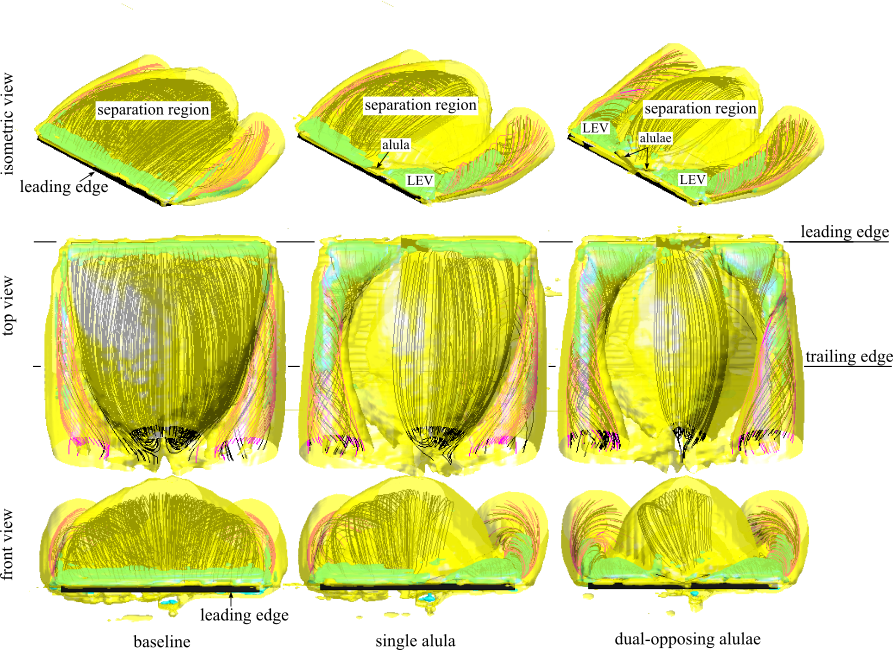}\\ 
   \end{minipage}     
        \vspace{-10pt}
       \caption{\footnotesize Time-averaged vortical flow over wing with no alula (\textit{left column}), single alula (\textit{middle column}), and dual-opposing alulae (\textit{right column}). Shown in yellow is the iso-surface of $||\boldsymbol{\omega}||_2 = 3$. Shown in cyan is the iso-surface of $Q = 13$ to indicate vortex cores. Streamlines stemming from the side-edge of the wing are colored magenta. Streamlines stemming from the leading-edge of the wing are colored black. Rows depict isometric views, top views (parallel to wing plane), and front views (orthogonal to wing plane), respectively.
}
     \vspace{-10pt}
   \label{fig:Vortiso}
\end{figure}

This section describes the three-dimensional time-averaged vortical flow over the wing with a single alula, and pair of alulae, in comparison to its plain wing counterpart. Fig. \ref{fig:Vortiso} plots two different isosurface values atop three-dimensional streamline patterns. Vortex sheets are identified by an isosurface of vorticity magnitude ($||\boldsymbol{\omega}||_2 = 3$) where vortex cores are identified by an isosurface of the Q-criterion (Q = 13). $Q$ locally measures the excess of rotation rate relative to strain rate (\cite{Moin:88b}) and for incompressible flows is given by $Q \equiv \frac{1}{2}(||\boldsymbol{\Omega}||^2 - ||\boldsymbol{S}||^2)$, where $\boldsymbol{\Omega} \equiv \frac{1}{2}[\boldsymbol{\nabla u} - (\boldsymbol{\nabla u})^T]$ and $\boldsymbol{S} \equiv \frac{1}{2}[\boldsymbol{\nabla u} + (\boldsymbol{\nabla u})^T]$. Streamlines are initialized at the leading edge of the wing at spacial locations of high-magnitude positive spanwise vorticity ($\omega_y > 15$). These leading-edge streamlines are colored black. Streamlines are also initiated at the side-edges of the wing and are uniformly spaced along the chord of the wing. These streamlines are colored magenta. 

Vortical flow over the baseline wing takes the form of blanketing shear layers that envelop a separation region. Streamlines stemming from the leading and side edges of the wing follow the trajectory of separated shear layers and do not reattach to the wing. We observe the separation region to extend downstream passed the wing's trailing-edge which implies that the flow over the baseline wing is `massively separated' in this time-averaged sense. 

The addition of a single alula to the wing results in two disparate vortical flow regions on the wing. For wing sections outboard of the alula, the leading-edge vortex sheet is pinned back to the wing plane. Underneath, leading-edge and wing-tip streamlines wind around the core of an aft-tilted leading-edge vortex (LEV) (as identified by Q). The LEV stems from the alula's root and sweeps across the outer corner of the wing merging with the tip flow to form an apparent recirculatory tip vortex by the trailing-edge of the wing. Adjacent wing sections remain topologically similar to that of the baseline wing as vortical flow takes the form of separated shear layers that blanket a separation region. However, the spanwise extent of the separation region is reduced. For the dual-opposing alula configuration, two LEV vortex systems span the outer edges of the left and right wing. In between these regions, lies a separation region of further reduced size in comparison to the single and baseline wing cases.

  %-----------------------------
\begin{figure}
\centering
     \vspace{0pt}
  \begin{minipage}{1\linewidth}
   \includegraphics[width=1\textwidth, angle=0]{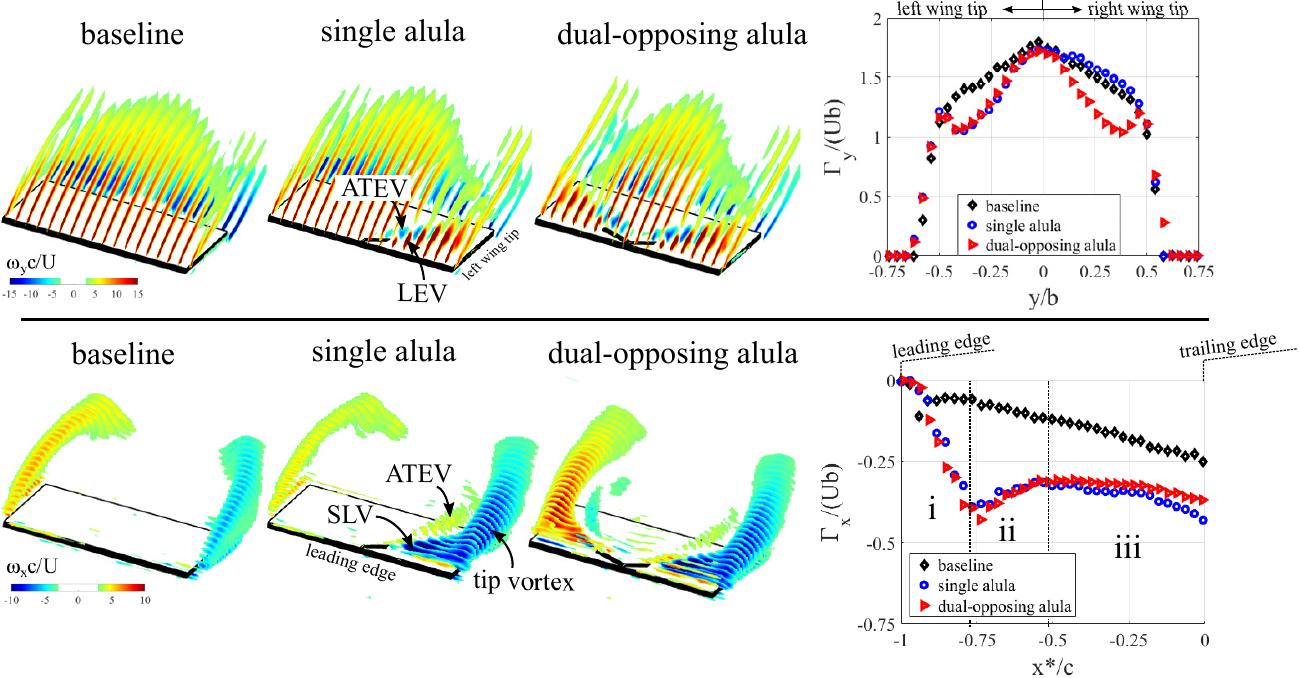}\\ 
   \end{minipage} 
        \vspace{-10pt}   
       \caption{\footnotesize Contour slices of time-averaged (\textit{top}) spanwise and (\textit{bottom}) streamwise vorticity for the wing affixed with no alula (baseline), single alula, and dual-opposing alulae. Plots on the right depict corresponding circulation distributions across the span and chord, respectively. Spanwise-oriented circulation is computed from positive spanwise vorticity in the flow. Streamwise-oriented circulation is computed from negative streamwise vorticity above the wing plane located on the left half of the wing. Regions marked i-iii are explained as follows: i) emergence and growth of the surface-layer vortex, SLV, via aft-tilting of LEV ii) amalgamation of SLV and side-edge shear layer into tip vortex, iii) tip vortex strengthening through feeding of side-edge shear layer.
}
     \vspace{-10pt}
   \label{fig:VortDistSlice}
\end{figure}

%----------------------------
For insight into the strength of vortices, we consider sectional cuts of spanwise- and streamwise-oriented vorticity, $\omega_y$ and $\omega_x$, respectively, in planes normal to the respective component of vorticity (see Fig. \ref{fig:VortDistSlice}). Accompanying these figures are plots comparing the corresponding circulation distributions across the span and chord, respectively. Spanwise-oriented circulation, $\Gamma_y$, is computed from $+\omega_y$, where streamwise-oriented circulation, $\Gamma_x$, is computed from $-\omega_x$ at $y/b < 0$ (left semispan of the wing). The distribution of $\Gamma_x$ is represented using a wing-aligned coordinate system where $x^*/c = -1$ marks the leading-edge of the wing, and, $x^*/c = 0$ marks the trailing edge of the wing.

  %-----------------------------
%----------------------------

The aft-tilted LEV has components in the $+\omega_y$ and $\mp \omega_x$ directions; the $\mp$ sign depending on if the alula is oriented to the left or right wing. Outboard of the alula $+\omega_y$ concentrates near the wing plane and lifts off as the wing tip is approached. $\omega_x$ comprising the LEV initiates as a surface layer (depicted as surface-layer vortex or `SLV') and with increasing downstream distance shifts outboard amalgamating with the side-edge shear layer, of like sign, forming a near-circular tip vortex by the midchord of the wing. Stemming from the trailing-edge of the alula lies a vortex, termed alula trailing-edge vortex or ATEV, that hugs the inner periphery of the LEV. This vortex is of opposite sign to the LEV containing vorticity of $-\omega_y$ and $\pm \omega_x$ direction; the $\pm$ sign depending on if the alula is oriented to the left or right wing. These results reflect the aft-tilted nature of vortices associated with the alula, i.e. the LEV and the ATEV. We will explain the tilting of these structures in section \ref{sec:tilt}.

From the circulation distributions shown in Fig. \ref{fig:VortDistSlice}, notable changes in spanwise and chordwise loading of the wing with an alula(e) are observed relative to the baseline wing. Spanwise stations harboring an LEV experience a reduction in $\Gamma_y$ which is approximately equal between the dual-opposing and single alula cases. This result suggests that the alula facilitates the removal and/or reorientation of $+\omega_y$ generated at the wing's leading-edge. In regard to streamwise-oriented circulation distributions, the emergence and growth of the SLV associated with the aft-tilted LEV results in the rapid increase in the magnitude of $-\Gamma_x$ which peaks around the chordwise location of $x^*/c = -0.75$. With increasing chordwise distance in the range $-0.75 < x^*/c < -0.5$, the amalgamation of the surface-layer vortex and side-edge shear layer into the tip vortex results in a reduction in magnitude of $-\Gamma_x$ due to the contraction of these vortices. From  $-0.25< x^*/c < 0$ the tip vortex strengthens as it is continues to be fed by the side-edge shear layer. We refer to the interested reader to the Fig. \ref{fig:VortVelX} in the Appendix which depicts the streamwise evolution of the SLV and its merging with the tip flow.

\section{LEV formation and maintenance}\label{sec:rollup}
Having described the vortical flow over the wing, we now investigate the mechanisms that enable the alula to induce and stabilize the observed attached vortex system. We start by considering the flow at spanwise stations in the vicinity of the alula to grant insight into the alula's interaction with the leading-edge flow and the formation of the LEV and oppositely-signed ATEV. A topological map is constructed to assist the discussion. Then we explain the mechanisms that drive the LEV to both tilt aft and smoothly merge with the tip flow. We then consider a series of tests performed at a higher angle of attack for which the effectiveness of the alula is hindered, and use our newfound knowledge of the aerodynamic mechanisms of the alula to slightly modify the alula in an attempt to restore its effectiveness at this high angle of attack condition.

\subsection{Topological description of LEV roll-up via alula}\label{sec:rollup}

\begin{figure}
\centering
     \vspace{0pt}
  \begin{minipage}{1\linewidth}
   \includegraphics[width=1\textwidth, angle=0]{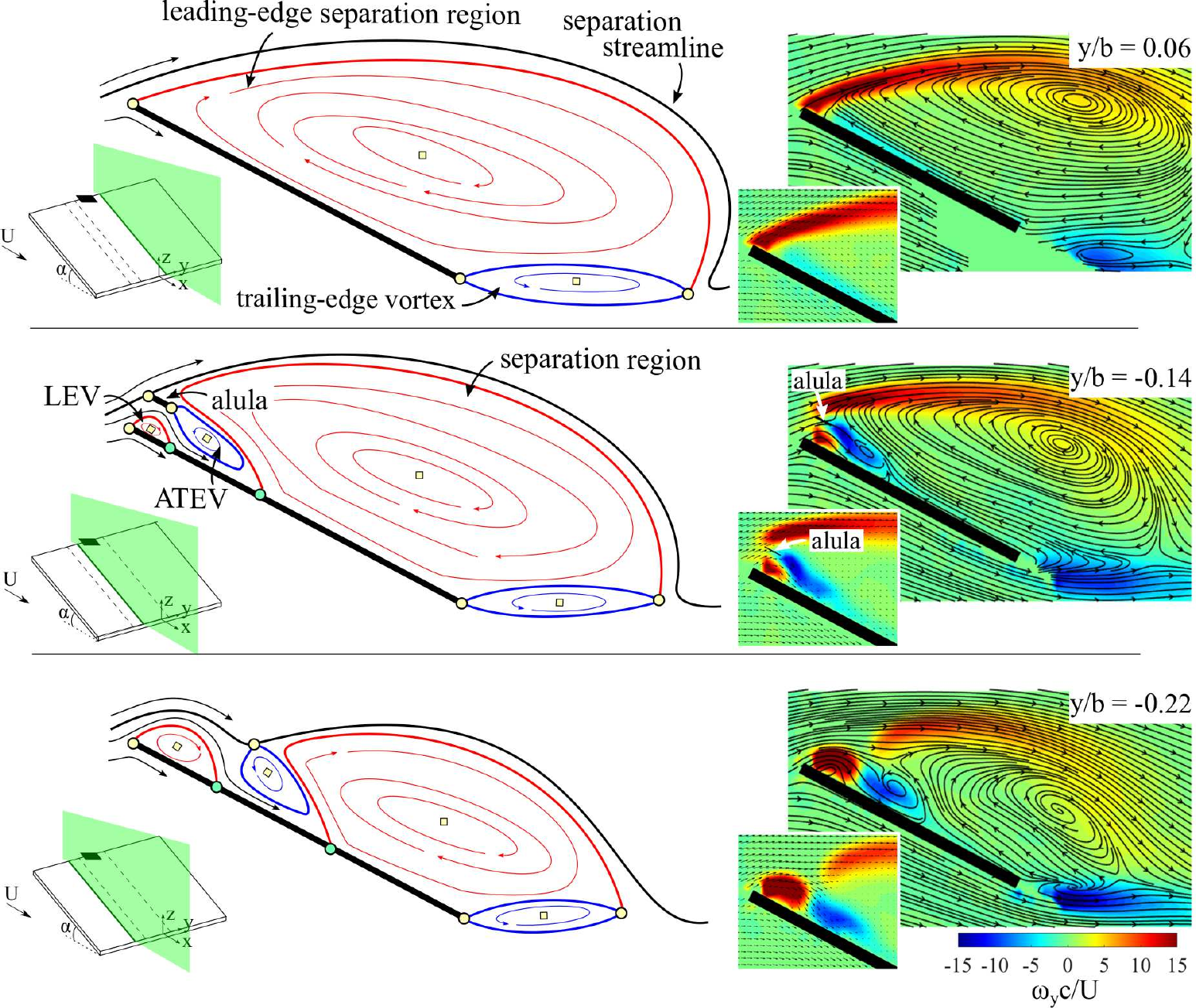}\\ 
   \end{minipage}     
        \vspace{-10pt}
       \caption{Topological description of the time-averaged flow over a wing with a single alula in streamwise planes at spanwise locations as indicated. Full-nodes and saddles are depicted by yellow squares and circles, respectively. Half-saddle points are depicted by cyan circles. Measured streamline patterns and spanwise vorticity contours are included on the right. Inset figures display measured velocity vectors near leading-edge of the wing.}
   \label{fig:Topology}
\end{figure}

We consider the case of the wing with a single alula and analyze streamline patterns and spanwise vorticity contours in chordwise planes in the immediate vicinity of the alula (refer to Fig. \ref{fig:Topology}). To assist the discussion, topological maps are created which have been previously used by the authors to describe the vortex shedding process on an airfoil (\cite{Mohseni:08l}) and the time-average mid-plane vortex structure on low-\AR\ wings at high incidences (\cite{Mohseni:17c}). From \cite{HuntJCR:78a} critical/fixed points in a two-dimensional plane section of the flow satisfy,
\begin{equation}
\big(\sum N + \frac{1}{2} \sum N'\big) - \big(\sum S + \frac{1}{2}\sum S'\big) = 1-n,
\label{eq:topo}
\end{equation}

\noindent where $S$ and $S'$ are full- and half-saddles, $N$ and $N'$ are full- and half-nodes and $n$ is the connectivity of the plane. $n = 3$ when the wing and alula are considered and $n = 2$ when the wing is considered.

  %-----------------------------
%----------------------------

We start by considering a plane of data, $y/b = 0.06$, located near the alula's root on the wing section opposite that for which the alula is overhung. Spanwise vorticity is organized into a shear layer stemming from the leading-edge of the wing which envelops a leading-edge separation region comprised of clockwise swirl. A trailing-edge vortex is also apparent comprised of anti-clockwise swirl. A topological flow description of the separated flow at the mid-plane of thin low-\AR\ wings was given by \cite{Mohseni:17c} which is adopted here. Two full-nodes represent the counter-rotating regions associated with the leading-edge separation region and the trailing-edge vortex, and three full-saddles represent the leading and trailing-edge stagnation points as well as the rear stagnation point associated with the leading-edge separation region. A separation streamline is added that blankets all recirculatory regions in the flow.

The topology of the flow on the wing sections outboard of the alula is fundamentally different than that on the adjacent wing section. The first plane considered ($y/b = -0.14$) is located immediately outboard of the alula, a distance $0.004b$ away from the alula's tip. The projection of the alula's tip on the contour plot of spanwise vorticity is as indicated. We observe the leading-edge shear layer to stem from the leading-edge of the alula as opposed to the leading-edge of the wing. Two additional vortices are introduced into the flow, the LEV and the ATEV, comprised of clockwise and anti-clockwise swirl. The LEV stems from the wing's leading-edge where the ATEV stems from the alula's trailing edge. Underneath the ATEV, lies high-magnitude flow near the surface of the wing as indicated by the magnitude of time-averaged velocity vectors there. This flow penetrates the separation region terminating at an apparent stagnation point as indicated by streamline patterns.

Streamlines in this plane of data are used to topologically describe the flow in a plane that slices through the alula and wing at a spanwise location coincident with the alula's wing tip. The topological description of this flow is explained as follows. With the addition of the alula, two bodies are now present, thus $n = 3$ in Eq. \ref{eq:topo}. Fixed points are the same as the topological description at $y/b =0.06$, however, two full saddles are added to represent the leading and trailing-edge stagnation points on the alula and two half saddles are added to represent additional stagnation points on the wing. Underneath the alula, the wing's leading-edge streamline impinges back on the wing surface forming a region of clockwise swirl (designated LEV). Aft of the LEV lies a streamline that both originates and terminates at the alula's trailing-edge. This streamline encircles a region of anti-clockwise swirl associated with the ATEV. With the LEV and ATEV occupying the leading-edge of the wing, the separation region is no longer enclosed by a streamline stemming from the leading-edge of the wing. Rather the streamline encircling the separation region stems from a half-saddle point on the wing located aft of the ALEV at approximately the 0.4c location. This streamline terminates at a full saddle point in the wake in a similar fashion to that which occurs at $y/b = 0.06$, however, this fixed point is closer to the trailing-edge of the wing resulting in a trailing-edge vortex, comprised of anti-clockwise swirl, of reduced size.

Lastly, we consider a plane of data outboard of the alula's tip at $y/b = -0.22$. The aft stagnation point of the LEV and the forward stagnation point of the separation region both move rearward, causing the enlargement of the LEV region and the shrinking of the separation region. In between, the ATEV, released from the alula, is suspended by a full saddle point formed from the merging of the full-saddle points located at the leading and trailing-edges of the alula. Curved streamlines associated with the enlarged LEV and reduced separation region indicate enhanced lift generation at spanwise stations outboard of the alula.

This analysis invites the current interpretation of the formation of the LEV and ATEV. The no-flow-through condition imposed by the alula, or the image vortex in the alula, steers separated leading-edge flow back to the wing plane. Shear produced from the penetration of this flow into the separation region results in the formation of the ATEV. The LEV forms from the accumulation of spanwise vorticity generated at the wing's leading-edge which is subsequently `trapped' by the oppositely-signed ATEV. While this pairing of the LEV and ATEV may assist in regulating LEV growth through vorticity annihilation it is unlikely that two-dimensional mechanisms alone are responsible for the attached LEV. As is discussed next, LEV growth appears to also be regulated by vortex tilting and outboard vorticity flux.

  %-----------------------------
\begin{figure}
\centering
     \vspace{0pt}
  \begin{minipage}{1\linewidth}
   \includegraphics[width=1\textwidth, angle=0]{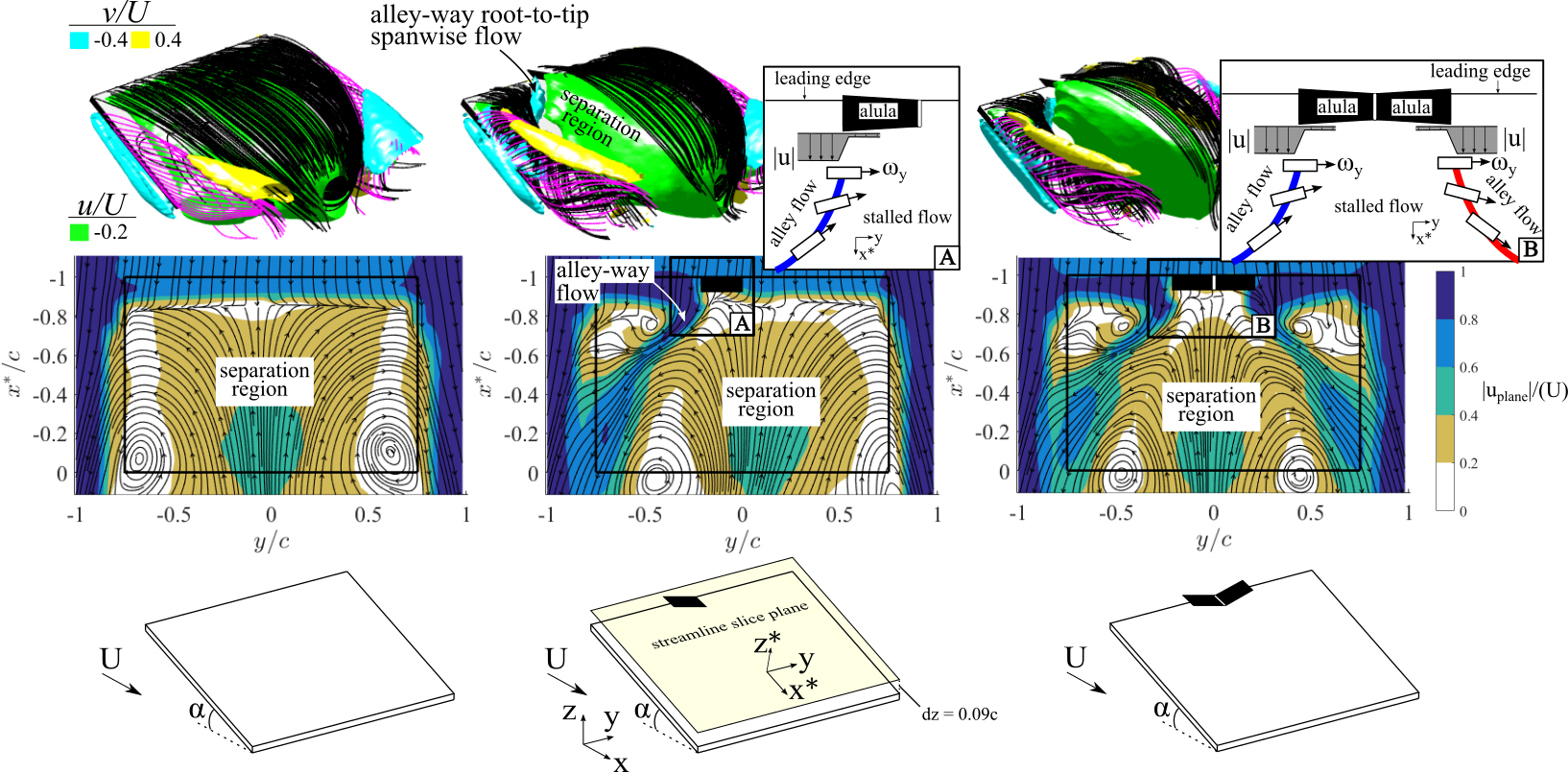}\\ 
   \end{minipage}    
        \vspace{-10pt} 
       \caption{\footnotesize Time-averaged velocity iso-surfaces for wing affixed with no alula (\textit{left column}), single alula (\textit{middle column}), and dual-opposing alulae (\textit{right column}). The two dimensional figures plot near-surface streamline patterns obtained at a vertical distance $dz = 0.09c$ from the top surface of the wing. Overlayed in these figures are contours of time-averaged in-plane velocity magnitude. Inset schematics depict the aft tilting of spanwise-oriented vorticity associated with boxed regions labeled A and B.}
     \vspace{-10pt}
   \label{fig:NearSurf}
\end{figure}
%----------------------------

\subsection{On the aft tilting of the LEV and outboard vorticity flux}\label{sec:tilt}
This subsection starts by describing how spanwise vorticity, steered back toward the wing plane by the alula, is subsequently tilted aft to form the SLV (recall Fig. \ref{fig:VortDistSlice}). This subsection ends by explaining the amalgamation of the SLV and side-edge shear layer into the tip vortex through an analysis of outboard vorticity flux.

To begin, we analyze near-surface streamline patterns and in-plane non-dimensional velocity magnitude contours in a plane parallel to the wing located a vertical distance $dz = 0.09c$ from the top surface of the wing (see Fig. \ref{fig:NearSurf}). Isometric views depicting three-dimensional streamline patterns and isosurfaces of the spanwise and streamwise-oriented velocity are also included for the readers convenience. From the near-surface streamline patterns, we observe high-magnitude root-to-tip flow ($>$$80\%$ that of the freestream velocity) to exist between two stagnation regions: one associated with the LEV that lies near the upstream corner of the wing and the other associated with the separation region which lies inboard of the alula's tip. A gradient in in-plane velocity magnitude is apparent immediately aft of the alula due the aforementioned high-magnitude `alley-way' flow lying adjacent to stalled flow in the separation region. As depicted in the schematics in Fig. \ref{fig:NearSurf}, a spanwise-aligned vortex element $+\omega_y$, generated at the wing's leading-edge and steered back to the wing plane by the alula, tilts aft when confronted with this gradient. The tilting of $+\omega_y$ on a leftward-oriented alula results in the production of $-\omega_x$ while the tilting of a rightward-oriented alula results in the production of $+\omega_x$. This is precisely the sign of vorticity comprising the SLV as observed in Fig. \ref{fig:VortDistSlice}.

  %-----------------------------
\begin{figure}
\centering
     \vspace{0pt}
  \begin{minipage}{1\linewidth}
   \includegraphics[width=1\textwidth, angle=0]{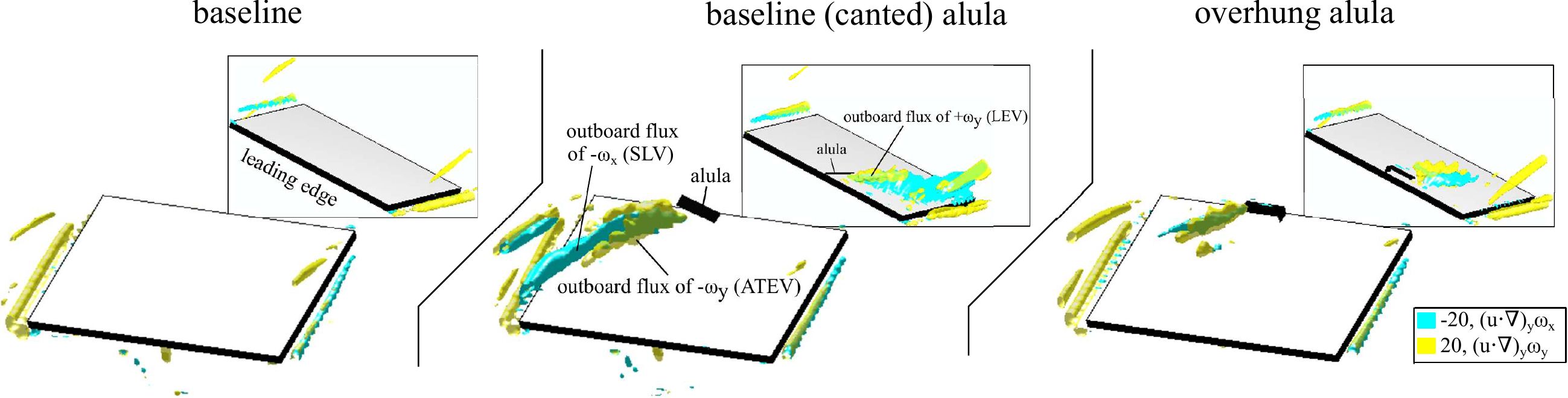}\\ 
   \end{minipage}     
        \vspace{-10pt}
       \caption{\footnotesize  Outboard convective flux of spanwise and streamwise vorticity shown by isosurfaces of $\boldsymbol{(u \cdot \nabla)_y}\omega_y$ and $\boldsymbol{(u \cdot \nabla)_y}\omega_x$ on baseline wing, wing with single canted alula, and wing with single overhung alula.}
     \vspace{-10pt}
   \label{fig:UgradOmega}
\end{figure}
%----------------------------

Having explained the likely mechanism behind the aft-tilted LEV, or SLV, next we investigate the mechanism that enables the merging of this vortex with the tip flow. In addition to facilitating vortex tilting, high magnitude root-to-tip spanwise flow produced by the alula drives an outboard flux of vorticity toward the wing tip. This is shown in Fig. \ref{fig:UgradOmega} which plots isosurfaces of $\boldsymbol{(u \cdot \nabla)_y}\omega_x = -20$ and $\boldsymbol{(u \cdot \nabla)_y}\omega_y = 20$ for the wing with a single canted alula in comparison to the baseline wing (the case of the overhung alula will be explained in the next subsection). On the wing with the canted alula, we observe an outboard flux of $+\omega_y$ and $-\omega_y$ which is associated with the transport of the non-tilted components of the LEV and ATEV, respectively. More importantly, we observe an outboard flux of $-\omega_x$ that emerges aft of the alula and is sustained across the outer wing section. This outboard flux of $-\omega_x$ drives the amalgamation of the SLV and the like-signed side-edge shear layer into a near circular tip vortex by the midchord of the wing. This is observed in contour plots of the baseline (canted) alula shown in Fig. \ref{fig:Amal} (see also Appendix Fig. \ref{fig:VortVelX}). These results emphasize the role of three-dimensional vorticity transport mechanisms (i.e. vortex tilting and outboard vorticity flux) in enabling the smooth merging of leading- and side-edge vortex flows and, in consequence, the attached LEV.

\subsection{On high-magnitude spanwise flow generation}\label{sec:spanflow}

We explained the role of the aft-located wall-jet of spanwise flow induced by the alula in facilitating the smooth merging of leading- and side-edge vortex flows. Still unclear is how the alula produces this flow. A starting point is to consider the orientation of the alula. The fact that the alula is canted and affixed to an inclined wing means that a component of the freestream is directed tangent to the alula. The tangential flow component is given as $U\sin{\alpha}\cos{\phi}$ where $U$ is the freestream velocity, $\alpha$ is the angle of attack, and $\phi$ is the alula's cant angle. For the test conditions considered, the freestream flow directed tangent to the alula is $0.18U$; a value substantially weaker than the spanwise velocity magnitudes measured which reach magnitudes $>0.80U$. Thus, there must be an additional mechanism driving the high-magnitude root-to-tip flow. 

\begin{figure}
\centering
     \vspace{0pt}
  \begin{minipage}{0.6\linewidth}
   \includegraphics[width=1\textwidth, angle=0]{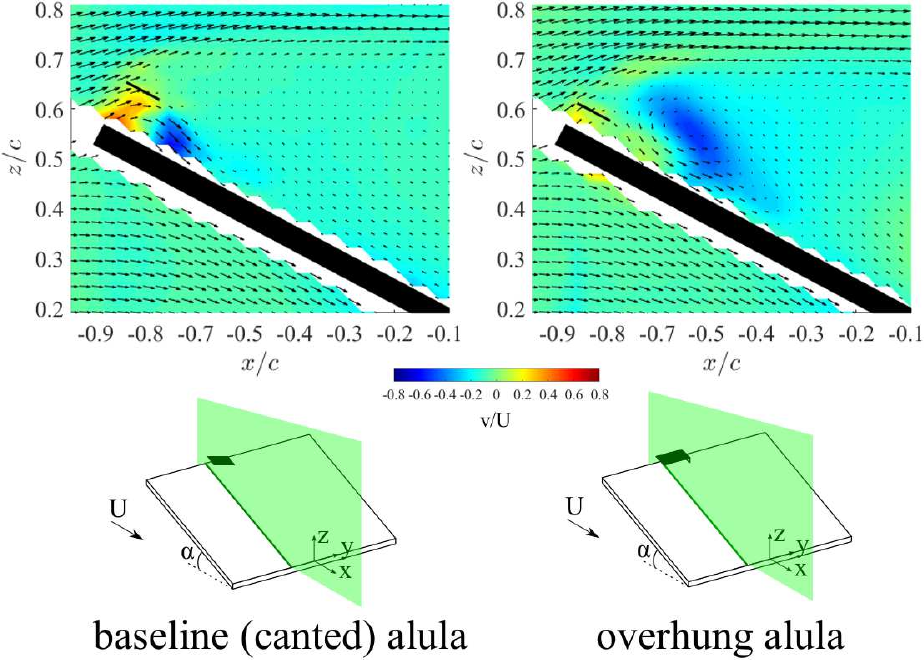}\\ 
   \end{minipage}    
        \vspace{-10pt} 
       \caption{\footnotesize Spanwise velocity contours in streamwise plane stationed at $y/b = -0.14$ for wing affixed with baseline (canted) alula (\textit{left column}) and overhung alula (\textit{right column}).}
     \vspace{-10pt}
   \label{fig:Spanflow}
\end{figure}

To assist this investigation, Fig. \ref{fig:Spanflow} plots spanwise velocity contours in streamwise planes at $y/b = -0.14$, a plane located immediately outboard of the alula. The projection of the alula's tip on this plane is plotted for reference. We observe an influx of spanwise flow at chordwise stations upstream of the alula's trailing-edge and an outboard flux of spanwise flow at chordwise stations downstream of the alula's trailing-edge. The inflow and outflow has a maximum spanwise velocity magnitude of $0.50U$ and $0.60U$, respectively. This organization of spanwise flow suggests that a region of clockwise swirl is established underneath the canted alula (when viewed from the top of the alula). Here, a fluid element stemming from the leading-edge of the wing in the vicinity of the alula is first directed toward the alula's root and, upon confronting it, is turned and subsequently ejected outboard toward the wing tip. The ability of the alula to entrain fluid not just from its front but also from its side would magnify the jet of flow directed outboard toward the wing tip by mass conservation which may explain the high magnitudes of root-to-tip spanwise flow produced by the alula.

We hypothesize the observed side-edge entrainment to be sustained by a spanwise pressure gradient associated with the canted alula. Here, the pressure gradient is the result of the contraction (and subsequent acceleration) experienced by spanwise flow directed toward the alula's root. We tested a modified alula which has no cant angle to corroborate this hypothesis. As opposed to being canted, the modified alula is overhung such as to form a small gap between its bottom surface and the wing's top surface. The gap between the `overhung alula' was set so that its frontal area was equal to that of the baseline (canted) alula. Moreover, the span of the overhung alula was kept the same as the projected span of the canted alula. 
  %-----------------------------
\begin{figure}
\centering
     \vspace{0pt}
  \begin{minipage}{1\linewidth}
   \includegraphics[width=1\textwidth, angle=0]{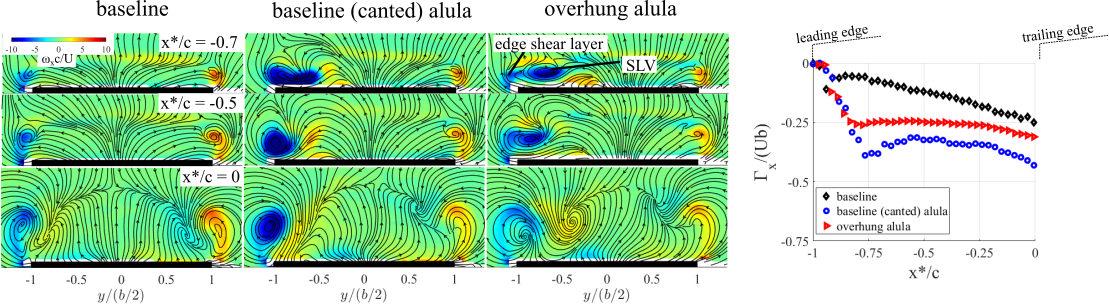}\\ 
   \end{minipage}     
        \vspace{-10pt}
       \caption{\footnotesize Time-averaged streamwise vorticity contours and streamlines in cross-stream planes at select chordwise locations. The plot on the right depicts the chordwise circulation distribution of the wing. Circulation is computed from negative streamwise vorticity above the wing plane located on the left half of the wing. }
     \vspace{-10pt}
   \label{fig:Amal}
\end{figure}
%----------------------------

Spanwise flow contours for the overhung alula are also plotted in Fig. \ref{fig:Spanflow}. The high magnitudes of inflow measured on the baseline (canted) alula are not observed on the overhung alula. Moreover, the outflow, or root-to-tip spanwise flow, produced by the overhung alula is weaker, more diffuse, and is displaced further aft of the trailing-edge of the alula in comparison to that which is observed on the baseline (canted) alula. This result lends itself to the notion that high-magnitude root-to-tip flow is connected to the cant angle of the alula. However, further research into high-magnitude spanwise flow generation via canted alula is necessary to resolve this connection.

The consequence of reduced spanwise flow generation is clear when comparing outboard vorticity flux (see Fig. \ref{fig:UgradOmega}) and chordwise circulation distributions  (see Fig. \ref{fig:Amal}) measured on the wing with the overhung alula compared to that of the baseline (canted) alula. From Fig. \ref{fig:UgradOmega}, the overhung alula, with reduced spanwise flow generation sees outboard vorticity flux localized near the alula. Without sustained vorticity flux toward the wing tip, the amalgamation process of the aft-tilted LEV, or SLV, with the edge shear layer is hindered resulting in a less coherent tip vortex by the trailing-edge of the wing as observed in Fig. \ref{fig:Amal}. This result is further reflected in chordwise circulation distributions of $\Gamma_x$ which depict consistently lower magnitudes of $-\Gamma_x$ for the overhung alula compared to the canted alula. These results emphasize the importance of spanwise flow generation on the stability of the vortex system exhibited by the wing with the alula and the apparent importance high-magnitude spanwise flow generation on enabling the smooth merging of leading- and side-edge flows and the apparent importance of the canted orientation of the alula.

  %-----------------------------
%----------------------------

\subsection{Shadowing of the alula at high angles of attack}\label{sec:carve}

\begin{figure}
\centering
     \vspace{0pt}
  \begin{minipage}{1\linewidth}
   \includegraphics[width=1\textwidth, angle=0]{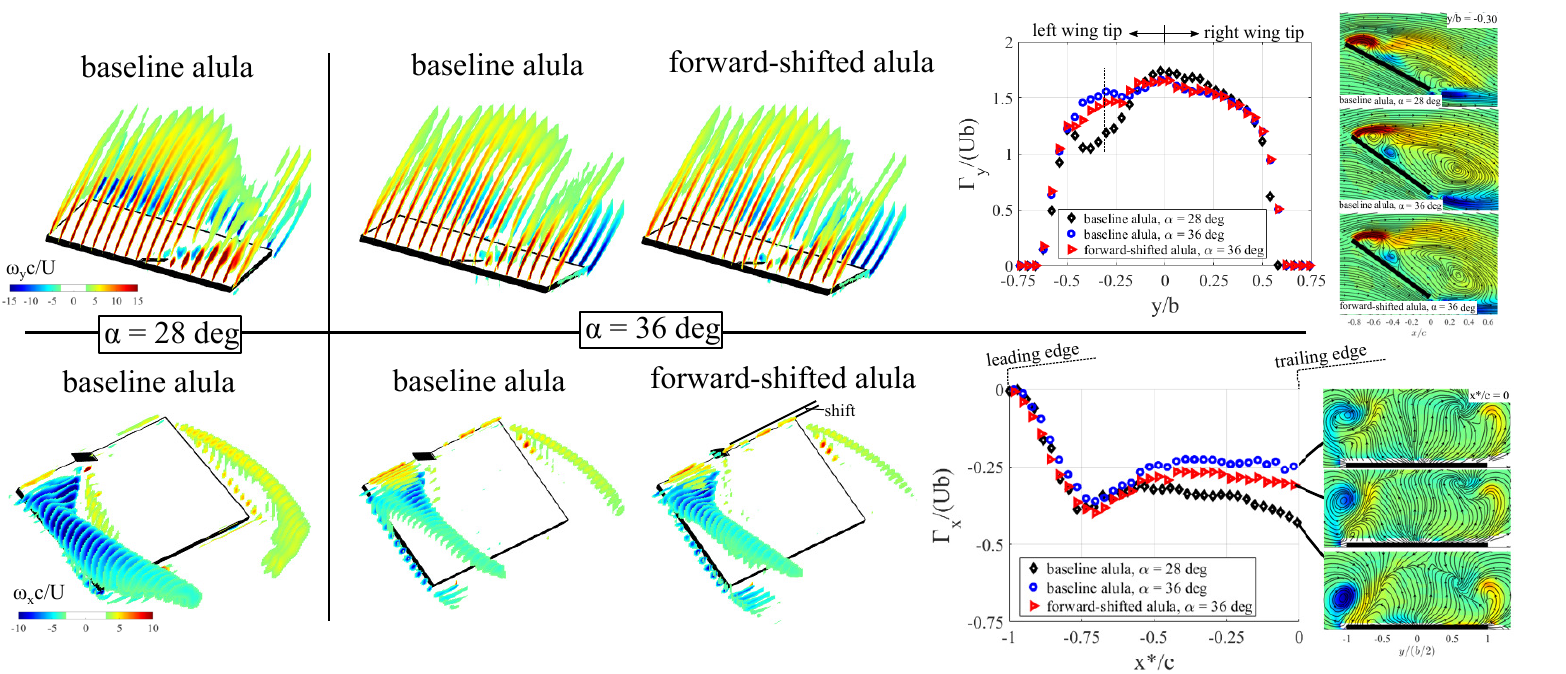}\\ 
   \end{minipage}     
        \vspace{-10pt}
       \caption{\footnotesize Contour slices of time-averaged non-dimensional (\textit{top}) spanwise and (\textit{bottom}) streamwise vorticity for the wing affixed with a single alula (baseline) and an alula that is shifted forward a distance one-half the chord length of the alula at angles of attack as marked. Plots depict corresponding circulation distributions across the span and chord respectively. Spanwise-oriented circulation is computed from positive spanwise vorticity in the flow. Streamwise-oriented circulation is computed from negative streamwise vorticity above the wing plane located on the left half of the wing. Contour plots at select spanwise and chord wise locations are also indicated.}
     \vspace{-10pt}
   \label{fig:Vortfshift}
\end{figure}

In the previous analysis, we connected LEV formation to the alula's ability to steer spanwise vorticity generated at the wing's leading-edge back to the wing plane. We showed this to manifest itself in distribution plots of spanwise circulation as a reduction in $\Gamma_y$ relative to the baseline wing with no alula. Recall, the reduction in $\Gamma_y$ was associated with spanwise vorticity in the LEV being reoriented and evacuated toward the wing tip via vortex tilting and outboard vorticity flux. The force measurements of \cite{Mohseni:19g} showed the alula to be ineffective if the angle of attack of the wing is too high. This suggests that the LEV formation and stabilization processes are hindered in some way. To elucidate this feature, we conducted experiments at a higher angle of attack $\alpha = 36$ deg. Spanwise and streamwise vorticity contours are compared in Fig. \ref{fig:Vortfshift} between the two angles of attack considered with spanwise- and chordwise- circulation distributions also plotted. At spanwise stations outboard of the alula, we observe $\Gamma_y$ to be significantly larger than at the higher angle of attack. Contour plots of spanwise circulation at $y/b = -0.3$ reveal spanwise vorticity to be more approximately organized into a leading-edge shear layer at the higher angle of attack as opposed to the concentrated LEV observed at the lower angle of attack. This inability to steer leading-edge vorticity back toward the wing plane to induce a recirculatory LEV means that vortex tilting and outboard vorticity flux via spanwise flow cannot take effect. The consequence is the downstream shedding of spanwise vortices into the wake manifesting itself as the diffuse shear layer observed in time averaged measurements. Moreover, with less spanwise vorticity being tilted, less streamwise vorticity is introduced into the flow resulting in a weaker and more diffuse tip vortex as is observed at the wing's trailing-edge of the wing, $x^*/c = 0$. 

We hypothesize that one of the reasons why the alula is less effective at the higher angle of attack is because the alula is less immersed in the wing's leading-edge flow due to it being shadowed behind the inclined wing. Because of this shadowing, the alula is less able to steer spanwise vorticity from the leading-edge shear layer back to the wing plane to induce the LEV. In an attempt to restore the alula's effectiveness at this higher angle of attack, we consider an identical alula that is shifted forward on the wing a distance one-half the chord length of the alula. Fig. \ref{fig:Vortfshift} includes contour plots of the forward-shifted alula with circulation measurements also plotted. We observe a reduction in $\Gamma_y$ on the outboard wing stations and subsequent increase in $-\Gamma_x$ on the aft portion of the wing; results that suggest that the forward-shifted alula, being less shadowed behind the wing, is more effective at steering leading-edge vorticity back to the wing plane. The ability to steer leading-edge vorticity back to the wing plane allows this flow to be tilted and subsequently evacuated toward the wing tip via high-magnitude spanwise flow generated by the canted alula.

  %-----------------------------
%----------------------------

\section{Concluding remarks}\label{sec:concl}
Based on the results described in this manuscript, we put forth the following explanation of the alula's maintenance of an attached aft-tilted leading-edge vortex (LEV). This flow pattern is the consequence of the alula's ability to smoothly merge otherwise separated leading and side-edge vortical flows. This is accomplished by leading-edge vorticity generated at the wing's leading-edge being steered back to the wing plane and subsequently subjected to an aft-located wall-jet of root-to-tip spanwise flow ($>80\%$ that of the freestream velocity). The latter feature induces LEV roll-up while the former feature tilts LEV vorticity aft and evacuates this flow toward the wing tip via an outboard vorticity flux. Our results further indicate that the hindrance of vortex steering, via the alula being shadowed behind the inclined wing, or the hindrance of high-magnitude root-to-tip spanwise flow generation, via elimination of the alula's cant angle, both result in a weakened and more diffuse vortex system (in the time-averaged sense). This leads us to conclude that it is the alula's ability to simultaneously i) steer leading-edge flow back to the wing plane and ii) subject this flow to a aft-located wall jet of high-magnitude root-to-tip spanwise flow that is critical to the alula's LEV stabilizing ability and thus enhanced wing lift at post-stall incidences.

The unique construction of the alula, which mimics a canted flap oriented toward the wing tip and stationed at the leading-edge of the wing, thus appears to possess all necessary ingredients to form and maintain an attached LEV on a steadily translating unswept wing that is inclined to the flow at post-stall angles. This capability of the alula would enable birds to maintain a spread-wing gliding posture to facilitate airbraking while enjoying enhanced wing lift via LEV flow to facilitate maneuvering. Additional work is necessary to the usefulness of an alula-inspired control effector for aircraft applications; although the need for such a device is apparent. Stealth aircraft require high-lift technologies with low observability whereas micro-aerial-vehicles require control effectors that are lightweight. The size of the alula, which in this study occupies an area 1\% that of the wing, makes an alula-inspired technology an attractive pain-reliever for these applications. However, key details such as the effects of maneuvering flight, leading-edge curvature of the wing, wing airfoil geometry, Reynolds number, among others on the attached vortex system associated with the alula still needs to be resolved.\\

\bibliography{RefA2,RefA2_Tom,RefA2_alula,RefA2_Control,RefA2_Validation,RefA2_actuator}
\bibliographystyle{jfm}

\section*{Acknowledgements}
The authors thank Dr. Adam DeVoria for discussions regarding this work which greatly improved the quality and clarity of this manuscript. The authors also acknowledge the partial financial support of the National Science Foundation (NSF). This work has resulted in the following patent: Sliding, Canted, Control Surfaces for Control Augmentation of Lifting Surfaces at High Angles of Attack,” U.S. Provisional Pat. Ser. No. 62/783,698, filed 12/21/2018. 

\section*{Declaration of Interests}
The authors report no conflict of interest.

\iftrue
\section{Appendix: Supplemental Representations of Three-Dimensional Post-Stall Flow}
The main text contains a curated analysis of the measured data which contributes to conclusions of the manuscript. However, researchers studying a similar problem may be interested in additional planar representations of the three-dimensional flow as is given here. 

\subsection{Contour plots of spanwise components of velocity and vorticity}
  %-----------------------------
\begin{figure}
\centering
     \vspace{0pt}
  \begin{minipage}{0.9\linewidth}
   \includegraphics[width=1\textwidth, angle=0]{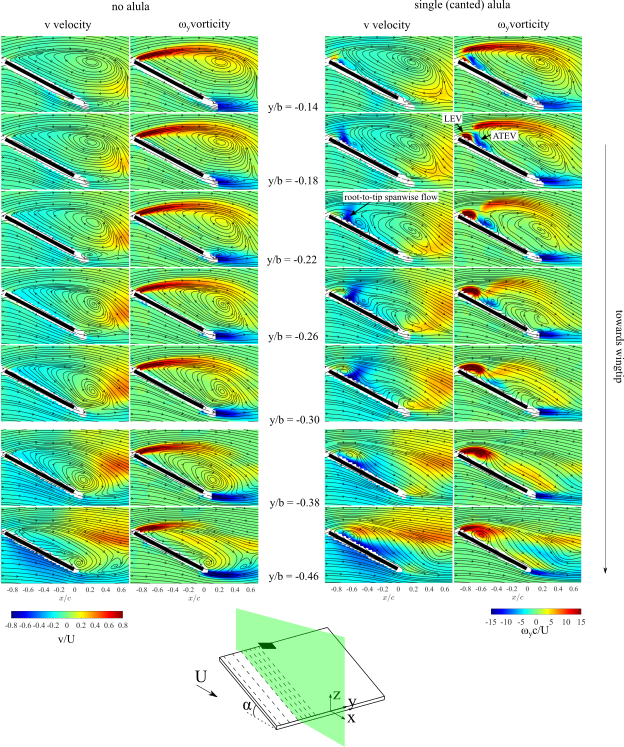}\\ 
   \end{minipage}     
        \vspace{-10pt}
       \caption{Time-averaged spanwise vorticity $\omega_yc/U$ and velocity $v/U$ contours in streamwise-oriented planes on the outer portion of the wing with (\textit{left column}) no alula and (\textit{right column}) a single leftward-oriented alula.}
   \label{fig:VortVelY}
           \vspace{-10pt}
\end{figure}
%----------------------------

Fig. \ref{fig:VortVelY} compares contour slices of spanwise velocity and spanwise vorticity in chordwise planes at different constant span values outboard of the wing with and without an alula. At spanwise stations immediately outboard of the alula, $-0.22 \leq  y/b \leq -0.14$, positive spanwise vorticity is organized into a concentrated leading-edge vortex (LEV) and a detached shear layer, the latter vortex pattern mimicking the separated leading-edge shear layer on the baseline wing. High-magnitude LEV flow penetrating the aft-located separation region results in the observed oppositely-signed vortex or ATEV. The strong trailing-edge vortex stemming from the wing is the consequence of mismatched flow at the trailing-edge of the wing due to the separation region extending into the wing wake.

At spanwise stations approaching the wing tip, $-0.22 \leq  y/b$, the chord-wise extent of the LEV grows and the ATEV is shifted downstream and becomes more diffuse. Subsequently, the separated region comprised of counter-clockwise rotating swirl is reduced in size with the enveloping shear layer reduced in magnitude. These results indicate that the LEV is conical due to its size increase as the tip is approached. Such growth of the LEV results in an a subsequent reduction in the chord-wise extent of the separation region near the wing tip. As the wing tip is approached, the trailing-edge vortex becomes less diffuse and is of lower magnitude than that measured on the wing without the alula. 

Spanwise velocity contours are also compared in Fig. \ref{fig:VortVelY}. High-magnitude spanwise flow is largely absent on the top surface of the baseline wing. In contrast, for the alula case, root-to-tip spanwise flow ($-v/U$) of magnitude $>$80\% that of the freestream velocity is measured on the wing with the alula. This flow takes the form of a tilted column positioned between the LEV and the aft ATEV. This column of root-to-tip spanwise flow then tilts in the downstream direction and flattens out as the tip is approached. In the main text, we show that this flow drives vortex tilting and outboard vorticity flux that curbs the downstream shedding of the LEV.

  %-----------------------------
\begin{figure}
\centering 
     \vspace{0pt}
  \begin{minipage}{1\linewidth}
   \includegraphics[width=1\textwidth, angle=0]{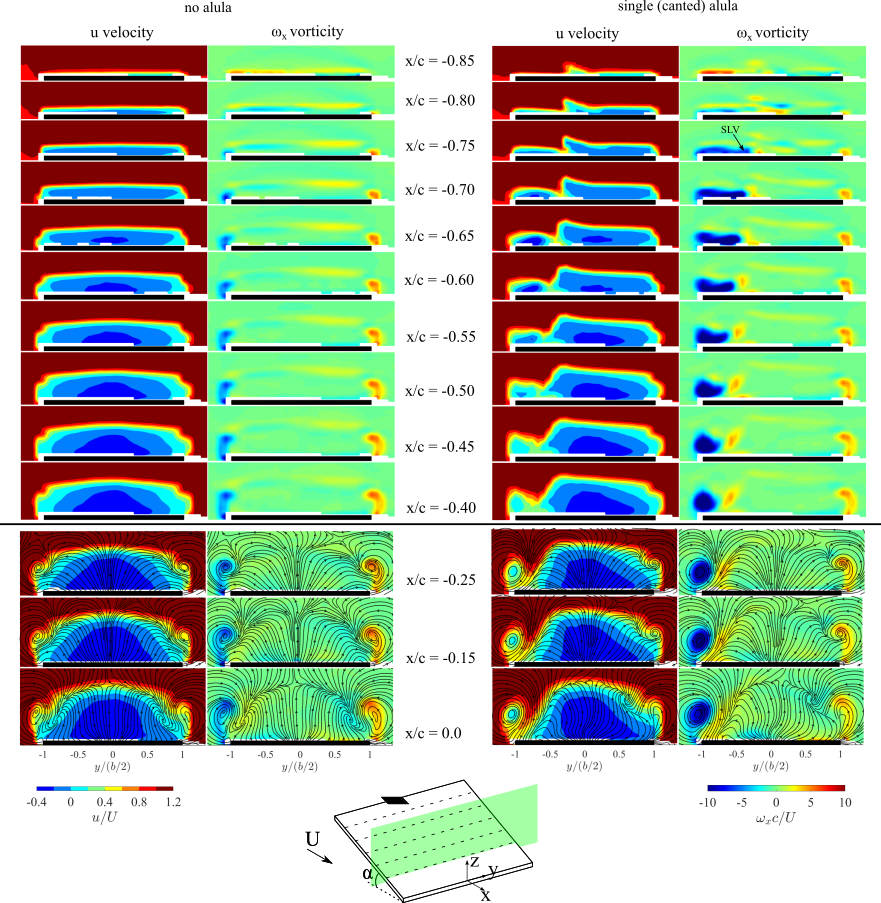}\\ 
   \end{minipage}     
           \vspace{-10pt}
       \caption{Time-averaged streamwise vorticity $\omega_xc/U$ and velocity $u/U$ contours in cross-stream planes on the outer portion of the wing with (\textit{left column}) no alula and (\textit{right column}) a single leftward-oriented alula.}
   \label{fig:VortVelX}
           \vspace{-10pt}
\end{figure}
%----------------------------

Coincident with positive spanwise vorticity in the LEV is tip-to-root spanwise flow of $~$20\% that of the freestream velocity. This organization of spanwise flow, consisting of tip-to-root spanwise flow lying adjacent to root-to-tip spanwise flow, shares some similarity to that observed on rotating wings (\cite{RinguetteMJ:13a}) and is the manifestation that the LEV is helical; a tracer particle injected at the leading-edge navigates the LEV by first moving inboard before being accelerated outboard upon reaching the aft end of the LEV.

\subsection{Contour plots of streamwise components of velocity and vorticity}

Streamwise vorticity contours are shown in Fig. \ref{fig:VortVelX}. For the baseline wing, streamwise vorticity takes the form of side-edge shear layers that roll up into weak concentric vortices at their ends. In contrast, for the alula cases, a surface layer vortex (or SLV) is observed near the upstream corners of the wing for wing sections outboard of the alula. This vorticity stems in part from the aft-tilted LEV. With downstream distance, this vorticity merges with the tip vortex, of like sign, forming a near circular tip vortex by the midchord of the wing.

\fi

\end{document}